\documentclass[11pt]{article}
\usepackage{amsmath,amssymb,amsfonts}
\usepackage[english]{babel}
\makeatletter
\@addtoreset{equation}{section}
\makeatother

\def\nn{\nonumber} \def\bd{\begin{document}} \def\ed{\end{document}}
\def\ds{\documentstyle}
\let\bm=\bibitem
\newcommand{\be}{\begin{equation}}
\newcommand{\ee}{\end{equation}}
\newcommand{\bea}{\setlength\arraycolsep{2pt} \begin{eqnarray}}
\newcommand{\eea}{\end{eqnarray}}
\newcommand{\hoch}[1]{$\, ^{#1}$}
\def\p{\partial}

\title{\large {\bf A new formula for conserved charges of
Lovelock gravity in AdS spacetimes and its generalization}}
\date{}

\author{Jun-Jin Peng$^{1,2}$\footnote{corresponding author: pengjjph@163.com},
\quad Hui-Fa Liu$^{1}$\footnote{hfaliu@163.com}\\ \\
\small \sl $^1$School of Physics and Electronic Science, Guizhou Normal University,\\
\small Guiyang, Guizhou 550001, People's Republic of China; \\
\small \sl  $^2$Guizhou Provincial Key Laboratory of Radio Astronomy and
Data Processing, \\
\small \sl Guizhou Normal University, \\
\small Guiyang, Guizhou 550001, People's Republic of China
}

\begin{document}

\maketitle
\vspace{20pt}

\begin{center}
\textbf{Abstract}
\end{center}

Within the framework of the Lovelock gravity theory, we propose a new
rank-four divergenceless tensor consisting of the Riemann curvature tensor
and inheriting its algebraic symmetry characters. Such a tensor can be
adopted to define conserved charges of the Lovelock gravity theory in
asymptotically anti-de Sitter (AdS) spacetimes. Besides, inspired with the
case of the Lovelock gravity, we put forward another general fourth-rank
tensor in the context of an arbitrary diffeomorphism invariant theory of
gravity described by the Lagrangian constructed out of the curvature
tensor. On basis of the newly-constructed tensor, we further suggest a
Komar-like formula for the conserved charges of this generic gravity
theory.

\voffset=-.90pt
\vspace{40pt}

\section{Introduction}\label{one}

As is well-known, Lovelock (or referred to as Lanczos-Lovelock) gravity
\cite{LoveGra} is the most natural higher-derivative extension of general
relativity to higher dimensions of spacetimes. Specifically, apart from the
cosmological constant, as well as the Einstein-Hilbert term,
the Lagrangians describing the Lovelock gravity theory involve quadratic
and higher-order polynomials of the curvature tensor with the degree relying
on the dimensions of spacetimes. In spite of this, the equations of motion
comprise merely up to second-order derivatives of the metric tensor. Such
behavior of the field equation enables
the Lovelock gravity to overcome the so-called Ostrogradsky instability,
which is a linear instability existing in the Hamiltonian associated with a
non-degenerate Lagrangian containing time derivative terms higher than the
first order \cite{Oinstab}.

In comparison with general
relativity, due to the existence
of the higher power curvature terms, the Lovelock gravity exhibits a series
of peculiar features. See for instance thermodynamical aspects in
references \cite{RievLL,PadGr,CLLth,DPPthG} and references therein. See for instance
geometrical aspects in references \cite{DaLov,BTZLov,DaPP} as well. As a consequence,
the Lovelock gravity has attracted an increasingly widespread attention
since 1990s. Particularly, some solutions have been found in works
\cite{ArEsreg,TBcstri,CFLP,DaLov,CaiOLov,BTZLov,DaPP,KofOl}. In order to interpret
these solutions, such as the first law of thermodynamics and other
thermodynamic properties,
an important question desired to address is to find proper
approaches to define the conserved charges of the Lovelock gravity. Till now,
some research has been devoted to this question from different perspectives
\cite{KofOl,FrHsur,KaSQLma,AHMMO,AHMMO2,PetLogr,BhMaj,ChaDa,GraCC,KasLGcc,DeBoSh,JakMy}
and several methods have been proposed. Despite all this, within the
present paper, motivated by the Abbott-Deser-Tekin (ADT) formalism
\cite{AbbottD,DeserT,DeserT2}
and the covariant phase space approach \cite{LeeWald,IyWald,WalZo} put forward by
Wald and his collaborators, we plan to provide another effective
formula to calculate the conserved charges of the Lovelock gravities in asymptotically anti-de Sitter (AdS) spacetime.
Apart from this, we attempt to extend it to generic diffeomorphism invariant theories
of gravity built out of the curvature tensor, with the Lagrangians given by
Eq. (\ref{CalLRiem}), that is, $\mathcal{L}_{Riem}=\sqrt{-g}L_{Riem}
\big(g^{\mu\nu},R_{\alpha\beta\rho\sigma}\big)$.

Quite recently, within the framework of the Einstein gravity theory described
by the Einstein-Hilbert-$\Lambda$ Lagrangian, a fourth-rank conserved tensor
that inherits the symmetries of the Riemann curvature tensor was suggested
in works \cite{PinGR1,PinGR}. By following the ADT
approach \cite{AbbottD,DeserT,DeserT2}, then it was adopted to define the
conserved charges of the Eistein gravity in asymptotically AdS spacetime.
In fact, if this conserved tensor is reformulated into the form involving
the generalized Kronecker delta, it will be demonstrated below that it can be
modified straightforwardly as a generic divergenceless tensor in the context
of the Lovelock gravity, which preserves the symmetries of the Riemann curvature
tensor as well. In terms of the perturbation of such a tensor about the AdS
background, we are going to propose another formula for the conserved charges of
the Lovelock gravity theory in asymptotically AdS spacetimes.

Subsequently, inspired with the specific example on the Lovelock gravity, we
shall go further and take into consideration of the definition for the
conserved charges of the diffeomorphism invariant theory of gravity,
described by the Lagrangian $\mathcal{L}_{Riem}$ in Eq. (\ref{CalLRiem}).
Our goal is to find a proper potential being of the similar form as the
Noether potential to define the conserved charges of such a gravity
theory in asymptotically AdS spacetimes. The resulted definition is required to
coincide with that via the ADT formalism or the covariant phase space method.
To achieve this, like in the case for the Lovelock gravity, we are going to
construct a general rank-four tensor inheriting the algebraic symmetry properties
of the Riemann curvature tensor. With help of the
newly-constructed tensor, then a 2-form Komar-like potential associated
with a Killing vector can be defined. The perturbation for
the potential on the AdS background further gives rise to a Komar-like
formula for the conserved charges of the general covariant gravity theory with
AdS asymptotics. By contrast, it will be showed that our formula for conserved
charges is successful in matching the one defined through the ADT method.

Throughout this paper, we will adopt the following notations and conventions.
We work in the case where the cosmological constant $\Lambda$ is negative
($\Lambda<0$). Nonetheless, we expect our results can be generalized to spacetimes
with a de Sitter ($\Lambda>0$) asymptotic. We make use of
units in which both the gravitational constant $G$ and the speed of light in
vacuum $c$ are set equal to one, namely, $G=c=1$. The positive integer $D$
denotes spacetime dimensions. The generalized Kronecker delta
$\delta^{\mu_1\cdot\cdot\cdot\mu_k}_{\nu_1\cdot\cdot\cdot\nu_k}$ is given by
$\delta^{\mu_1\cdot\cdot\cdot\mu_k}_{\nu_1\cdot\cdot\cdot\nu_k}=
k!\delta^{[\mu_1}_{[\nu_1}\cdot\cdot\cdot\delta^{\mu_k]}_{\nu_k]}$. For instance,
when $k=2$, $\delta^{\mu\nu}_{\rho\sigma}
=2\delta^{[\mu}_\rho\delta^{\nu]}_\sigma
=\delta^\mu_\rho\delta^\nu_\sigma
-\delta^\nu_\rho\delta^\mu_\sigma$. The objects with a bar such as the
covariant derivative $\bar{\nabla}$ or the metric tensor $\bar{g}_{\mu\nu}$
represent the ones defined with respect to the AdS background metric
$\bar{g}_{\mu\nu}$. For example,
$\bar{R}_{\mu\nu\rho\sigma}=R_{\mu\nu\rho\sigma}(g_{\alpha\beta}
\rightarrow\bar{g}_{\alpha\beta})$, and
$\bar{P}^{\mu\nu\rho\sigma}_{(0)}=2\bar{g}^{\rho[\mu}\bar{g}^{\nu]\sigma}$
is the counterpart of the rank-four tensor
$P^{\mu\nu\rho\sigma}_{(0)}=2g^{\rho[\mu}g^{\nu]\sigma}$ on the AdS
background.

The layout of this paper goes as follows. In section \ref{two}, a new
rank-four conserved tensor will be constructed and utilized to define
the conserved charges of the Lovelock gravity theories in the AdS spacetime.
Section \ref{three}
is devoted to the comparison between the conserved charges built from that
newly-constructed tensor and the ones via the ADT formalism. In section
\ref{four}, motivated by the case of the Lovelock gravity, we are going to
put forward another two generic tensors to give a definition of the
conserved charges for general covariant gravity theories built out of the
curvature tensor.

\section{The divergenceless tensor $\mathcal{P}^{\mu\nu\rho\sigma}_{(i)}$
and the conserved charges of the Lovelock gravities}\label{two}
In this section, we are going to propose a new rank-4 divergence-free tensor
$\mathcal{P}^{\mu\nu\rho\sigma}_{(i)}$, which depends on the Riemann curvature
tensor and preserves its symmetries. Then such a tensor will be applied to
define the conserved charges of various solutions with the AdS asymptotic in
the framework of Lovelock gravities.

Let us start with the $D$-dimensional Einstein gravity described by the well-known
Einstein-Hilbert-$\Lambda$ Lagrangian
\be
\mathcal{L}_{EH}=\sqrt{-g}(R-2\Lambda)
=\frac{1}{2}\sqrt{-g}\big(R^{\rho\sigma}_{~~\mu\nu}\delta^{\mu\nu}_{\rho\sigma}
-4\Lambda\big)
\, . \label{EHLagden}
\ee
Accordingly, the action for the Einstein gravity is of the form
$S_{EH}=(16\pi)^{-1}\int\mathcal{L}_{EH}dx^D$.
The derivative of the scalar $L_{EH}=\mathcal{L}_{EH}/\sqrt{-g}$ with respect
to the Riemann tensor gives rise to
\be
P^{\mu\nu}_{(0)\rho\sigma}=
2\frac{\partial L_{EH}}{\partial R^{\rho\sigma}_{~~~\mu\nu}}
=\delta^{\mu\nu}_{\rho\sigma}
\, , \label{P0def}
\ee
or $P^{\mu\nu\rho\sigma}_{(0)}=2g^{\rho[\mu}g^{\nu]\sigma}$.
It is easy to verify that the tensor $P_{(0)}^{\mu\nu\rho\sigma}$ possesses
the same index symmetries as the Riemann tensor and it is divergence-free,
namely, $\nabla_\mu P_{(0)}^{\mu\nu\rho\sigma}=0$. In works
\cite{PinGR1,PinGR}, by making use of $P_{(0)\rho\sigma}^{\mu\nu}$,
together with a divergence-free tenor $P^{\mu\nu}_{(1)\rho\sigma}$, given by
\bea
P^{\mu\nu}_{(1)\rho\sigma}&=&
R^{\mu\nu}_{~~\rho\sigma}
-4R^{[\mu}_{[\rho} \delta^{\nu]}_{\sigma]}
+\frac{1}{2} R \delta^{\mu\nu}_{\rho\sigma} \nn \\
&=&\frac{1}{4}R^{\alpha\beta}_{~~\gamma\lambda}
\delta^{\gamma\lambda\mu\nu}_{\alpha\beta\rho\sigma}
\, , \label{EP1def}
\eea
the authors proposed a new conserved tensor
\be
\mathcal{P}^{\mu\nu\rho\sigma}=P^{\mu\nu\rho\sigma}_{(1)}
-\frac{(D-2)(D-3)\ell}{2} P^{\mu\nu\rho\sigma}_{(0)}
\, , \label{CalPdef}
\ee
to define the conserved charges of the Einstein gravity theory in
asymptotically AdS spacetimes by following the ordinary ADT approach
\cite{AbbottD,DeserT,DeserT2}. In Eq. (\ref{CalPdef}) and what follows,
the constant parameter $\ell$ is related to the cosmological constant
$\Lambda$ through the relation,
\be
\ell=\frac{2\Lambda}{(D-1)(D-2)}
\, . \label{elldef}
\ee
Actually, motivated by the form of $P^{\mu\nu}_{(1)\rho\sigma}$ expressed
by the last equality in Eq. (\ref{CalPdef}), the conserved rank-4 tensor
$\mathcal{P}^{\mu\nu\rho\sigma}$ can be naturally generalized to the Lovelock
gravity theory. This will be explicitly demonstrated in the remainder of this section.

Now, we consider the Lovelock-type Lagrangians $\mathcal{L}^{(i)}_L$
$(0\leq i\leq [D/2-1])$ in a $D$-dimensional spacetime with the metric
$g_{\mu\nu}$, taking the form \cite{LoveGra}
\bea
\mathcal{L}^{(i)}_L&=&\sqrt{-g}\Big(L^{(i)}_L-2\tilde{\Lambda}\Big) \, , \nn \\
L^{(i)}_L&=&\frac{1}{4^i(i+1)}
\delta^{\gamma_1\lambda_1\cdot\cdot\cdot\gamma_{i+1}\lambda_{i+1}}
_{\alpha_1\beta_1\cdot\cdot\cdot\alpha_{i+1}\beta_{i+1}}
\prod_{r=1}^{i+1}R^{\alpha_r\beta_r}_{~~~~\gamma_r\lambda_r}
\, , \label{LovGLag}
\eea
where the constant parameter $\tilde{\Lambda}$ depends on the cosmological
constant $\Lambda$, since it is assumed that the Lagrangian
$\mathcal{L}^{(i)}_L$ allows the existence of (asymptotically) AdS solutions.
Particularly, if $i=0$ and $\tilde{\Lambda}=2\Lambda$,
$\mathcal{L}^{(0)}_L=2\mathcal{L}_{EH}$. Through the
derivative of the scalar $L^{(i)}_L$ with respect to the Riemann tensor,
we are able to define a fourth-rank tensor $P^{\mu\nu\rho\sigma}_{(i)}$ as
\cite{PadGr}
\bea
P^{\mu\nu}_{(i)\rho\sigma}&=&
\frac{\partial L^{(i)}_L}{\partial R^{\rho\sigma}_{~~~\mu\nu}} \nn \\
&=&\frac{1}{4^i}R^{\alpha_1\beta_1}_{~~~~\gamma_1\lambda_1}\cdot\cdot\cdot
R^{\alpha_i\beta_i}_{~~~~\gamma_i\lambda_i}
\delta^{\gamma_1\lambda_1\cdot\cdot\cdot\gamma_i\lambda_i\mu\nu}
_{\alpha_1\beta_1\cdot\cdot\cdot\alpha_i\beta_i\rho\sigma}
\, . \label{PRidef}
\eea
One can check that $P^{\mu\nu\rho\sigma}_{(i)}$ inherits the following
symmetries of the Riemann curvature tensor:
\bea
P^{\mu\nu\rho\sigma}_{(i)}&=&-P^{\nu\mu\rho\sigma}_{(i)}
=-P^{\mu\nu\sigma\rho}_{(i)} \, , \nn \\
P^{\mu\nu\rho\sigma}_{(i)}&=&P^{\rho\sigma\mu\nu}_{(i)}\, , \quad
P^{[\mu\nu\rho]\sigma}_{(i)}=0
\, . \label{Pisymme}
\eea
Apart from this, it satisfies the divergence-free equation
$\nabla_\mu P^{\mu\nu\rho\sigma}_{(i)}=0=\nabla_\nu P^{\mu\nu\rho\sigma}_{(i)}$,
as well as
$\nabla_\rho P^{\mu\nu\rho\sigma}_{(i)}=0=\nabla_\sigma P^{\mu\nu\rho\sigma}_{(i)}$,
arising from the fact that
\bea
\nabla_\mu P^{\mu\nu}_{(i)\rho\sigma}
&=&\frac{i}{4^i}\nabla_{[\mu}R^{\alpha_1\beta_1}_{~~~~\gamma_1\lambda_1]}
\cdot\cdot\cdot R^{\alpha_i\beta_i}_{~~~~\gamma_i\lambda_i}
\delta^{\gamma_1\lambda_1\cdot\cdot\cdot\gamma_i\lambda_i\mu\nu}
_{\alpha_1\beta_1\cdot\cdot\cdot\alpha_i\beta_i\rho\sigma} \nn \\
&=&0 \, . \nn
\eea
To obtain the second equality, we have made use of the Bianchi identity
$\nabla_{[\lambda}R^{\mu\nu}_{~~\rho\sigma]}=0$. For more information on the tensor
$P^{\mu\nu\rho\sigma}_{(i)}$, for instance, see the Refs. \cite{DaLLBia,KaRieL}.
By letting $P^{\mu\nu\rho\sigma}_{(i)}$ subtract a term proportional to the
tensor $P^{\mu\nu\rho\sigma}_{(0)}$ rather than to its counterpart
$\bar{P}^{\mu\nu\rho\sigma}_{(0)}$ on the AdS spacetime, where
$\bar{P}^{\mu\nu\rho\sigma}_{(0)}=2\bar{g}^{\rho[\mu}\bar{g}^{\nu]\sigma}$
with the help of the metric tensor $\bar{g}^{\mu\nu}$ for the AdS spacetime,
we put forward a divergenceless rank-four conserved tensor
$\mathcal{P}^{\mu\nu\rho\sigma}_{(i)}$, being of the form
\be
\mathcal{P}^{\mu\nu\rho\sigma}_{(i)}=
P^{\mu\nu\rho\sigma}_{(i)}-\lambda_{(i)}P^{\mu\nu\rho\sigma}_{(0)}
\, , \quad
\lambda_{(i)}=\Big(\frac{\ell}{2}\Big)^i\frac{(D-2)!}{(D-2i-2)!}
\, . \label{MCPidef}
\ee
For the $D$-dimensional AdS spacetime endowed with the Riemann curvature
tensor
\be
\bar{R}^{\mu\nu}_{~~~\rho\sigma}=\ell \delta^{\mu\nu}_{\rho\sigma}
\, , \label{AdSRT}
\ee
it is observed that $\bar{P}^{\mu\nu\rho\sigma}_{(i)}=
P^{\mu\nu\rho\sigma}_{(i)}\big(g_{\alpha\beta}\rightarrow
\bar{g}_{\alpha\beta}\big)
=\lambda_{(i)}\bar{P}^{\mu\nu\rho\sigma}_{(0)}$, giving rise to that
$\bar{\mathcal{P}}^{\mu\nu\rho\sigma}_{(i)}=
\mathcal{P}^{\mu\nu\rho\sigma}_{(i)}
\big(g_{\alpha\beta}\rightarrow\bar{g}_{\alpha\beta}\big)$ identically
vanishes. This plays a key role in the construction of the formula for the
conserved charges. Particularly, when
$i=1$, $\mathcal{P}^{\mu\nu\rho\sigma}_{(1)}$ becomes the tensor
$\mathcal{P}^{\mu\nu\rho\sigma}$ given by Eq. (\ref{CalPdef}). To
understand more on the tensor $\mathcal{P}^{\mu\nu\rho\sigma}_{(i)}$,
some remarks will be made in the next section.

Next, assumed that $\xi^\mu$ is a Killing vector of the spacetime with the metric $g_{\mu\nu}$,
we follow the covariant phase space method \cite{LeeWald,IyWald,WalZo} to
propose a 2-form superpotential $\mathcal{K}^{\mu\nu}_{(i)}$ as
\bea
\mathcal{K}^{\mu\nu}_{(i)}&=&
\mathcal{P}^{\mu\nu\rho\sigma}_{(i)}\nabla_\rho\xi_\sigma
=K^{\mu\nu}_{(i)}
-2\Big(\frac{\ell}{2}\Big)^i\frac{(D-2)!}{(D-2i-2)!}\nabla^{[\mu}\xi^{\nu]}
\, , \nn \\
K^{\mu\nu}_{(i)}&=&P^{\mu\nu\rho\sigma}_{(i)}\nabla_\rho\xi_\sigma
\, , \label{defCalKi}
\eea
where the skew-symmetric tensor $K^{\mu\nu}_{(i)}$ is the Noether potential.
Perturbating $\mathcal{K}^{\mu\nu}_{(i+1)}$ on the AdS spacetime,
we arrive at
\bea
\delta\mathcal{K}^{\mu\nu}_{(i+1)}&=&
\big(\delta\mathcal{P}^{\mu\nu}_{(i+1)\rho\sigma}\big)
\bar{\nabla}^\rho\bar{\xi}^\sigma
+\bar{\mathcal{P}}^{\mu\nu}_{(i+1)\rho\sigma}
\delta\big(\nabla^\rho\xi^\sigma\big)
\nn \\
&=&\big(\delta\mathcal{P}^{\mu\nu}_{(i+1)\rho\sigma}\big)
\bar{\nabla}^\rho\bar{\xi}^\sigma\nn \\
&=&\frac{(i+1)\ell^i}{2^i}\frac{(D-4)!}{(D-2i-4)!}
\delta\mathcal{K}^{\mu\nu}_{(1)}
\, , \label{delCalKi}
\eea
where $\bar{\xi}^\sigma$ is demanded to be the Killing vector of the AdS
background $\bar{g}_{\mu\nu}$. To obtain the second equality in
Eq. (\ref{delCalKi}), we have used
$\bar{\mathcal{P}}^{\mu\nu}_{(i+1)\rho\sigma}=0$.
When $i=1$, the quantity
$\delta\mathcal{K}^{\mu\nu}_{(1)}$ can be written as \cite{PinGR1,PinGR}
\bea
\delta\mathcal{K}^{\mu\nu}_{(1)}
&=&
\big(\delta\mathcal{P}^{\mu\nu}_{(1)\rho\sigma}\big)
\bar{\nabla}^\rho\bar{\xi}^\sigma \nn \\
&=&-2\ell(D-3) \bar{Q}^{\mu\nu}_{EH}+\bar{\nabla}_\gamma
\bar{U}^{\gamma\mu\nu}_{(1)}
\, , \label{delCalP1}
\eea
with the 2-form $\bar{Q}^{\mu\nu}_{EH}$ given by \cite{AbbottD,DeserT,DeserT2}
\be
\bar{Q}^{\mu\nu}_{EH}=\frac{1}{2}\bar{P}^{\mu\nu}_{(0)\rho\sigma}
\Big(h^{[\rho}_\lambda\bar{\nabla}^{\sigma]}\bar{\xi}^\lambda
+\bar{\xi}^\lambda\bar{\nabla}^{[\rho} h^{\sigma]}_\lambda\Big)
+\frac{1}{4}h\bar{P}^{\mu\nu}_{(0)\rho\sigma}\bar{\nabla}^\rho\bar{\xi}^\sigma
-\bar{\xi}^{[\mu}\bar{P}^{\nu]\lambda}_{(0)~\rho\sigma}
\bar{\nabla}^\sigma h^\rho_\lambda
\, , \label{QforGR}
\ee
as well as the 3-form $\bar{U}^{\gamma\mu\nu}_{(1)}$ taking the form
\be
\bar{U}^{\gamma\mu\nu}_{(1)}=\frac{1}{2}
\delta^{\lambda\gamma\mu\nu}_{\alpha\beta\rho\sigma}
\big(\bar{\nabla}^\alpha h^\beta_\lambda\big)
\bar{\nabla}^\rho\bar{\xi}^\sigma
\, . \label{Ubar1def}
\ee
In Eqs. (\ref{QforGR}) and (\ref{Ubar1def}),
$\bar{P}^{\mu\nu}_{(0)\rho\sigma}=\delta^{\mu\nu}_{\rho\sigma}$ and
the perturbation
$h_{\mu\nu}=\delta g_{\mu\nu}$ of the metric $g_{\mu\nu}$ is defined through
\be
h_{\mu\nu}=g_{\mu\nu}-\bar{g}_{\mu\nu}
\, . \label{hmndef}
\ee
We shall see that the 2-form $\bar{Q}^{\mu\nu}_{EH}$ is just the ADT potential
of the Einstein gravity theory described by the Lagrangian (\ref{EHLagden})
in the following section.

Finally, supposed that there exists a subregion given by a $(D-1)$-dimensional
hypersurface $\Sigma$ with the boundary $\partial\Sigma$,\footnote{To match
the ordinary AdS background metric, it is reasonable to set up the coordinate system
$\{x^\mu\}=\{t,r,x^i\}$ $(i=1,2,\cdot\cdot\cdot,D-2)$, where $r$ is the radial
coordinate. Thus, as usual, $\partial\Sigma$ denotes the surface with $t=const$ and
$r=\infty$ when the conserved charge $\mathcal{Q}_{(i)}$ represents the mass
or the angular momentum. As an example, see the explicit calculations on the
conserved charges of black holes in the works \cite{JJPEPJC,PJJGCC}.} we are
able to make use of the 2-form $\delta\mathcal{K}^{\mu\nu}_{(i+1)}$ to define
the conserved charge $\mathcal{Q}_{(i)}$ associated with
the Lovelock gravity with Lagrangian (\ref{LovGLag}) in this subregion.
The surface charge $\mathcal{Q}_{(i)}$ takes the following form
\bea
\mathcal{Q}_{(i)}&=&-\frac{1}{8\pi(i+1)(D-2i-3)\ell}\int_{\partial\Sigma}
\delta\mathcal{K}^{\mu\nu}_{(i+1)} d\Sigma_{\mu\nu} \nn \\
&=&-\frac{\ell^{i-1}}{2^{i+3}\pi}\frac{(D-4)!}{(D-2i-3)!}
\int_{\partial\Sigma} \delta\mathcal{K}^{\mu\nu}_{(1)} d\Sigma_{\mu\nu} \nn \\
&=&\frac{\ell^i}{2^{i-1}}\frac{(D-3)!}{(D-2i-3)!}\mathcal{Q}_{EH}
\, , \label{CCinCalK}
\eea
where the conserved charge $\mathcal{Q}_{EH}$ can be regarded as
the ADT one for the Einstein gravity theory
\cite{BhMaj,DeserT,KimKY,JJPEPJC}, presented by
\be
\mathcal{Q}_{EH}=
\frac{1}{8\pi}\int_{\partial\Sigma} \bar{Q}^{\mu\nu}_{EH} d\Sigma_{\mu\nu}
\, . \label{CCofEinGr}
\ee
What is more, let us take into consideration of the conserved charges for the full
Lovelock Lagrangian $\mathcal{L}_L$, having the form
\be
\mathcal{L}_L=\sqrt{-g}\sum_{i=0}^{[D/2-1]}c_i L^{(i)}_L
-2\Lambda\sqrt{-g}
\, , \label{FulLLag}
\ee
where $c_i$'s are arbitrary coupling constants in principle. In terms of the formula
(\ref{CCinCalK}), the definition $\mathcal{Q}_L$ for the conserved charges
associated with the Lagrangian (\ref{FulLLag}) in asymptotically AdS
spacetime can be presented by
\bea
\mathcal{Q}_L&=&\sum_{i=0}^{[D/2-1]}c_i \mathcal{Q}_{(i)} \nn \\
&=&\mathcal{Q}_{EH}\sum_{i=0}^{[D/2-1]}
\frac{c_i\ell^i}{2^{i-1}}\frac{(D-3)!}{(D-2i-3)!}
\, , \label{CCofFulLL}
\eea
which coincides with the corresponding result recently obtained through the
so-called field-theoretical approach in \cite{PetLogr}.
As a significant application, the formula (\ref{CCinCalK}) can be utilized to compute
the conserved charges of solutions in the Lovelock gravity theories, for
instance, the ones in the references
\cite{ArEsreg,TBcstri,CFLP,DaLov,CaiOLov,BTZLov,DaPP,KofOl}.
Besides, inspired with the works \cite{JMen,PFLbh,MCGS}, the formula
(\ref{CCinCalK}) may be modified properly to derive the entropy and the first
law of the black hole solutions, particularly in the framework of the Lovelock
gravities formulated in terms of orthonormal coframes.

What is more, for the Lovelock gravity theory described by the
Lagrangian (\ref{FulLLag}), apart from
the ordinary AdS spacetime, this theory can admit maximally-symmetric AdS
spacetimes with the Riemann curvature tensor $R^{\mu\nu}_{\rho\sigma}
=-\ell_{eff}^{-2}\delta^{\mu\nu}_{\rho\sigma}$, where the effective AdS
radius $\ell_{eff}$, solved from the field equation, depends on the
coupling constants $c_i$'s \cite{AHMMO,AHMMO2}. Particularly, under some
conditions with respect to the coupling constants, there exist degenerate
AdS spacetimes on which the linear perturbation to the expression for the
equation of motion disappears. As a consequence, when such spacetimes are
chosen as the reference background, the ADT approach yields vanishing
charges. This is attributed to the fact that the conserved ADT current,
proportional to the linear perturbation of the expression for the equation
of motion, vanishes identically. The formula (\ref{CCofFulLL}) encounters
the same outcome since we shall see below that the potentials  involved in
this formula coincide with the ones via the ADT approach. Regarding to this,
we conclude that the ADT approach, as well as the method in this paper,
breaks down to the Lovelock theories with a vacuum degeneracy. By contrast,
in the works \cite{AHMMO,AHMMO2}, a strategy has been worked out to derive a
conserved charge expression for the asymptotically AdS solutions having
degenerate vacua in the Lovelock gravities. The resulted charges are
dependent of the degeneracy of the theories.

\section{Comparison with the ADT formalism}\label{three}
In the present section, the formula (\ref{CCinCalK}) for the conserved charges
will be compared with the one via the ADT formalism
\cite{AbbottD,DeserT,DeserT2}.

In terms of the works \cite{BhMaj,KimKY,JJPEPJC}, the perturbation on an arbitrary
background gives rise to the (off-shell) ADT potential
$Q_{(i)}^{\mu\nu}$ associated to the Lovelock Lagrangian (\ref{LovGLag}), being
of the form
\bea
Q_{(i)}^{\mu\nu}&=&\frac{1}{\sqrt{-g}}
\delta\Big(\sqrt{-g}K^{\mu\nu}_{(i)}\Big)-\xi^{[\mu}\Theta^{\nu]}_{(i)}
\nn \\
&=&\delta\big(P^{\mu\nu\rho\sigma}_{(i)}
\nabla_\rho\xi_\sigma\big)+\frac{1}{2}g^{\alpha\beta}\big(\delta
g_{\alpha\beta}\big)
P^{\mu\nu\rho\sigma}_{(i)}\nabla_\rho\xi_\sigma \nn \\
&&-2\xi^{[\mu}P^{\nu]\lambda\rho\sigma}_{(i)}
\nabla_\sigma\delta g_{\rho\lambda}
\, . \label{ADTpoten}
\eea
Here the surface term $\Theta^{\mu}_{(i)}=2P^{\mu\nu\rho\sigma}_{(i)}
\nabla_\sigma\delta g_{\rho\nu}$, and the variation of the Noether potential
$K^{\mu\nu}_{(i)}=P^{\mu\nu\rho\sigma}_{(i)}\nabla_\rho\xi_\sigma$ can be
further expressed as
\bea
\delta K^{\mu\nu}_{(i)}&=&-\frac{2i}{4^i}
\xi^\omega\big(\nabla^\alpha h^\beta_\lambda\big)
\delta^{\gamma_1\lambda_1\cdot\cdot\cdot\gamma_{i-1}\lambda_{i-1}\gamma\lambda\mu\nu}
_{\alpha_1\beta_1\cdot\cdot\cdot\alpha_{i-1}\beta_{i-1}\alpha\beta\rho\sigma}
R^{\rho\sigma}_{\gamma\omega}
\prod_{r=1}^{i-1}R^{\alpha_r\beta_r}_{\gamma_r\lambda_r}\nn \\
&&-\frac{i}{4^i}h^\beta_\theta\big(\nabla^\rho\xi^\sigma\big)
\delta^{\gamma_1\lambda_1\cdot\cdot\cdot\gamma_{i-1}\lambda_{i-1}\gamma\lambda\mu\nu}
_{\alpha_1\beta_1\cdot\cdot\cdot\alpha_{i-1}\beta_{i-1}\alpha\beta\rho\sigma}
R^{\alpha\theta}_{\gamma\lambda}
\prod_{r=1}^{i-1}R^{\alpha_r\beta_r}_{\gamma_r\lambda_r}\nn \\
&&+P^{\mu\nu}_{(i)\rho\sigma}\delta\big(\nabla^\rho\xi^\sigma\big)
+\nabla_\gamma U^{\gamma\mu\nu}_{(i)}
\, , \label{delPixi}
\eea
with the 3-form $U^{\gamma\mu\nu}_{(i)}$ given by
\be
U^{\gamma\mu\nu}_{(i)}=
\frac{2i}{4^i}\big(\nabla^\alpha h^\beta_\lambda\big)
\big(\nabla^\rho\xi^\sigma\big)
\delta^{\gamma_1\lambda_1\cdot\cdot\cdot\gamma_{i-1}\lambda_{i-1}\lambda\gamma\mu\nu}
_{\alpha_1\beta_1\cdot\cdot\cdot\alpha_{i-1}\beta_{i-1}\alpha\beta\rho\sigma}
\prod_{r=1}^{i-1}R^{\alpha_r\beta_r}_{\gamma_r\lambda_r}
\, . \label{ThreFUidef}
\ee
In Eqs. (\ref{delPixi}) and (\ref{ThreFUidef}), $h^\mu_\nu$ stands for that
$h^\mu_\nu=g^{\mu\rho}\delta g_{\rho\nu}$. It is worth noting that
the ADT potential $Q_{(i)}^{\mu\nu}$ can be also obtained via the well-known
covariant phase space approach \cite{LeeWald,IyWald,WalZo} put forward by
Wald and his collaborators
(for example, see the works \cite{RievLL,PadGr,FrHsur,KaSQLma}). In particular,
taking into account the $i=1$ case of Eq. (\ref{delPixi}), one obtains
\bea
\delta\big(P^{\mu\nu\rho\sigma}_{(1)}
\nabla_\rho\xi_\sigma\big)&=&
P^{\mu\nu}_{(1)\rho\sigma}\delta\big(\nabla^\rho\xi^\sigma\big)
-\frac{1}{4}h^\beta_\theta R^{\alpha\theta}_{\gamma\lambda}
\delta^{\gamma\lambda\mu\nu}_{\alpha\beta\rho\sigma}
\nabla^\rho\xi^\sigma \nn \\
&&+\frac{1}{2}R^{\rho\sigma}_{\gamma\omega}
\delta^{\gamma\lambda\mu\nu}_{\alpha\beta\rho\sigma}\xi^\omega
\nabla^\beta h^\alpha_\lambda
+\frac{1}{2}\nabla_\gamma
\Big[\delta^{\lambda\gamma\mu\nu}_{\alpha\beta\rho\sigma}
\Big(\nabla^\alpha h^\beta_\lambda\Big)
\nabla^\rho\xi^\sigma\Big]
\, . \label{delP1xi}
\eea
This equation naturally yields the 2-form $\delta\mathcal{K}^{\mu\nu}_{(1)}$
in Eq. (\ref{delCalP1}) when the reference background is the AdS metric
$\bar{g}_{\mu\nu}$.

Let us pay attention to the ADT potential $Q_{(i)}^{\mu\nu}$ on the fixed AdS
reference background with the Riemann curvature tensor
(\ref{AdSRT}). In such a case, $Q_{(i)}^{\mu\nu}$ turns into
\be
\bar{Q}_{(i)}^{\mu\nu}=
\Big(\frac{\ell}{2}\Big)^{i-1}\frac{(D-4)!}{(D-2i-2)!}\Big[
(D-3)(D-2i-2)\ell\bar{Q}^{\mu\nu}_{EH}
+i\bar{\nabla}_\gamma \bar{U}^{\gamma\mu\nu}_{(1)}\Big]
\, . \label{Qibarmunu}
\ee
The above equation leads to that
$\bar{Q}^{\mu\nu}_{EH}=\bar{Q}_{(0)}^{\mu\nu}/2$, verifying the statement
in the previous section that $\bar{Q}^{\mu\nu}_{EH}$ is the ADT potential
for the Einstein gravity theory. According to Eqs. (\ref{delCalKi}) and
(\ref{Qibarmunu}), the relation between $\bar{Q}_{(i)}^{\mu\nu}$ and
$\delta\mathcal{K}^{\mu\nu}_{(i+1)}$ is read off as
\be
\bar{Q}_{(i)}^{\mu\nu}=
\frac{\ell^{i-1}}{2^i(D-3)}\frac{(D-2)!}{(D-2i-2)!}
\bar{\nabla}_\gamma \bar{U}^{\gamma\mu\nu}_{(1)}
-\frac{\delta\mathcal{K}^{\mu\nu}_{(i+1)}}{(i+1)(D-2i-3)\ell}
\, . \label{ReQiKip1}
\ee
This demonstrates that the ADT potential $\bar{Q}_{(i)}^{\mu\nu}$ is equivalent with
the superpotential $\delta\mathcal{K}^{\mu\nu}_{(i+1)}$. Therefore, in the
framework of the Lovelock gravity, one can conclude that the ADT formula is
consistent with the formula (\ref{CCinCalK}). Furthermore, the equivalence
between $\bar{Q}_{(i)}^{\mu\nu}$ and $\delta\mathcal{K}^{\mu\nu}_{(i+1)}$
demonstrates that the subtracted term related to
$P^{\mu\nu\rho\sigma}_{(0)}$ in Eq. (\ref{MCPidef}) compensates the contribution
from the surface term $\Theta^\mu_{(i)}$.

Inspired with the relationship
between $\bar{Q}_{(i)}^{\mu\nu}$ and $\delta\mathcal{K}^{\mu\nu}_{(i+1)}$,
we make some remarks on the divergence-free tensor $\mathcal{P}^{\mu\nu\rho\sigma}_{(i)}$
given by Eq. (\ref{MCPidef}). Firstly, we emphasize that the substracted
term $2\lambda_{(i)}g^{\rho[\mu}g^{\nu]\sigma}$ in $\mathcal{P}^{\mu\nu\rho\sigma}_{(i)}$ can not be naively replaced
with the one $\bar{P}^{\mu\nu\rho\sigma}_{(i)}
=2\lambda_{(i)}\bar{g}^{\rho[\mu}\bar{g}^{\nu]\sigma}$ associated
to the AdS background metric, that is to say, the tensor
$P^{\mu\nu\rho\sigma}_{(0)}$ in
$\mathcal{P}^{\mu\nu\rho\sigma}_{(i)}$ can not be changed as
its counterpart $\bar{P}^{\mu\nu\rho\sigma}_{(0)}$ upon the reference
background. Otherwise,
$\delta\big(\bar{P}^{\mu\nu\rho\sigma}_{(0)}\nabla_\rho\xi_\sigma\big)
=\delta\big(P^{\mu\nu\rho\sigma}_{(0)}\nabla_\rho\xi_\sigma\big)
+4h^{\lambda[\mu}\bar{\nabla}_\lambda\bar{\xi}^{\nu]}$ brings
an additional subtracted term
$4\lambda_{(i)}h^{\lambda[\mu}\bar{\nabla}_\lambda\bar{\xi}^{\nu]}$
to the potential $\delta\mathcal{K}^{\mu\nu}_{(i+1)}$,
rendering it inconsistent with $\bar{Q}_{(i)}^{\mu\nu}$. Secondly, the
tensor $\mathcal{P}^{\mu\nu\rho\sigma}_{(i)}$
is not uniquely qualified to the construction of the potential
$\mathcal{K}^{\mu\nu}_{(i)}$ in Eq. (\ref{defCalKi}). As a matter of fact,
through the linear combination, the tensor
$\mathcal{P}^{\mu\nu\rho\sigma}_{(i)}$ can be generalized as\footnote{We
thank the anonymous referee for pointing out this.}
\be
\check{\mathcal{P}}^{\mu\nu\rho\sigma}_{(i+1)}=\sum_{j=0}^{i}
\beta_j\mathcal{P}^{\mu\nu\rho\sigma}_{(j+1)}
\, , \label{chcaPi}
\ee
with the constant parameters $\beta_j$'s constrained by
\be
\sum_{j=0}^{i}\beta_j\frac{(j+1)\ell^j}{2^j}\frac{(D-4)!}{(D-2j-4)!}
=\frac{(i+1)\ell^i}{2^i}\frac{(D-4)!}{(D-2i-4)!}
\, .
\ee
On basis of $\check{\mathcal{P}}^{\mu\nu\rho\sigma}_{(i+1)}$, the potential
$\mathcal{K}^{\mu\nu}_{(i)}$ is reexpressed as
$\mathcal{K}^{\mu\nu}_{(i+1)}=\check{\mathcal{P}}^{\mu\nu\rho\sigma}_{(i+1)}
\nabla_\rho\xi_\sigma$. In spite of this, in order to guarantee that the
definition of the conserved charges matches the one via the ADT formalism
and the covariant phase space method in a straightforward and simple manner,
then it appears natural to adopt $\mathcal{P}^{\mu\nu\rho\sigma}_{(i)}$
rather than its linear combinations to define the conserved charges. Thirdly,
the vanishing of $\bar{\mathcal{P}}^{\mu\nu\rho\sigma}_{(i)}$
yields $\bar{\mathcal{K}}^{\mu\nu}_{(i)}=0$, further giving rise to that
\be
\delta\mathcal{K}^{\mu\nu}_{(i)}=
\mathcal{K}^{\mu\nu}_{(i)}-\bar{\mathcal{K}}^{\mu\nu}_{(i)}
=\mathcal{K}^{\mu\nu}_{(i)}
\, . \nn
\ee
As a consequence, the perturbation of the
potential $\delta\mathcal{K}^{\mu\nu}_{(i+1)}$, involved in the formula
(\ref{CCinCalK}) for the conserved charges, can be replaced by
$\mathcal{K}^{\mu\nu}_{(i+1)}$. To this point, one is able to see that
the formula (\ref{CCinCalK}) for the conserved charges of the Lovelock gravity
theory has the advantage of avoiding the perturbation for the potential.
Thus it is much simpler than the one via the ADT approach.

\section{The formula of conserved charges for generic diffeomorphism
invariant theories of gravity}
\label{four}

In this section, inspired with the tensor $\mathcal{P}^{\mu\nu\rho\sigma}_{(i)}$,
we are going to put forward another two generic rank-4 tensors and make use
of each of them to define the conserved charges of an arbitrary gravity
theory that is constructed from the curvature tensor in asymptotically
AdS spacetimes.

We take into consideration of the general gravity theory described by
the Lagrangian $\mathcal{L}_{Riem}=\sqrt{-g}L_{Riem}$, where the scalar
$L_{Riem}$ is treated as a functional built from the
Riemann curvature tensor $R_{\alpha\beta\rho\sigma}$ in combination with
the metric tensor $g^{\mu\nu}$, that is,
\be
\mathcal{L}_{Riem}=\sqrt{-g}L_{Riem}
\big(g^{\mu\nu},R_{\alpha\beta\rho\sigma}\big)
\, . \label{CalLRiem}
\ee
Apparently, $\mathcal{L}_{Riem}$ covers the Lovelock-type Lagrangians.
Like in \cite{RievLL}, through the derivative of the scalar
$L_{Riem}$ with respect to the Riemann tensor $R_{\mu\nu\rho\sigma}$, we
are able to define a tensor $P_R^{\mu\nu\rho\sigma}$ as
\be
P_R^{\mu\nu\rho\sigma}=\frac{\partial L_{Riem}}{\partial R_{\mu\nu\rho\sigma}}
\, . \label{PR4def}
\ee
$P_R^{\mu\nu\rho\sigma}$ exhibits the same symmetries as the Riemann tensor, namely,
$P_R^{\mu\nu\rho\sigma}=-P_R^{\nu\mu\rho\sigma}=-P_R^{\mu\nu\sigma\rho}$ and
$P_R^{\mu\nu\rho\sigma}=P_R^{\rho\sigma\mu\nu}$. Nevertheless,
unlike the conserved tensor $P_{(i)}^{\mu\nu\rho\sigma}$,
here the tensor $P_R^{\mu\nu\rho\sigma}$ could be divergence-free or not.
Besides, it is supposed that the value of $P_R^{\mu\nu\rho\sigma}$ on the
background AdS spacetime fulfils
\be
\bar{P}^{\mu\nu}_{R~\rho\sigma}=
P^{\mu\nu}_{R~\rho\sigma}\big(g_{\alpha\beta}\rightarrow\bar{g}_{\alpha\beta}\big)
=k\delta^{\mu\nu}_{\rho\sigma}
\, , \label{barPR4def}
\ee
where $k$ is a certain constant parameter. It should be pointed out that
the condition (\ref{barPR4def}) can be always guaranteed. This is attributed
to the fact that $P^{\mu\nu\rho\sigma}_{R}$ can be expressed as the linear
combination of the terms having the general form
\be
\big(g^{\bullet\bullet}\cdot\cdot\cdot g^{\bullet\bullet}
R_{\bullet\bullet\bullet\bullet}\cdot\cdot\cdot
R_{\bullet\bullet\bullet\bullet}\big)^{\alpha\beta\gamma\lambda}
\delta^{\mu\nu}_{\alpha\beta}\delta^{\rho\sigma}_{\gamma\lambda}
+\big([\mu\nu]\leftrightarrow[\rho\sigma]\big)
\, . \label{PRgeform}
\ee
As a consequence, when $g^{\mu\nu}=\bar{g}^{\mu\nu}$ and
$R_{\mu\nu\rho\sigma}=2\ell\bar{g}_{\mu[\rho}\bar{g}_{\sigma]\nu}$, the above
expression $(\cdot\cdot\cdot)^{\mu\nu}_{\rho\sigma}$ must be proportional
to the generalized Kronecker delta $\delta^{\mu\nu}_{\rho\sigma}$. With
the help of the rank-4 tensor $P_R^{\mu\nu\rho\sigma}$, we put forward
another fourth-rank tensor $\mathcal{P}_R^{\mu\nu\rho\sigma}$, being
of the form
\be
\mathcal{P}^{\mu\nu}_{R~\rho\sigma}=
P^{\mu\nu}_{R~\rho\sigma}-k\delta^{\mu\nu}_{\rho\sigma}
-\frac{k}{(D-3)\ell} \mathcal{P}^{\mu\nu}_{(1)\rho\sigma}
\, . \label{CalPRdef}
\ee
By contrast with the tensor $\mathcal{P}^{\mu\nu}_{(i)\rho\sigma}$, obviously,
here
$\bar{\mathcal{P}}^{\mu\nu}_{R~\rho\sigma}=\mathcal{P}^{\mu\nu}_{R~\rho\sigma}
\big(g_{\alpha\beta}\rightarrow\bar{g}_{\alpha\beta}\big)$ still identically disappears,
and $\mathcal{P}^{\mu\nu\rho\sigma}_{R}$ inherits the symmetries of the Riemann
curvature tensor as well.
In particular, for the Einstein gravity,
$\mathcal{P}^{\mu\nu}_{R~\rho\sigma}=
-[2(D-3)\ell]^{-1}\mathcal{P}^{\alpha\beta}_{~~\rho\sigma}$
since $P^{\mu\nu}_{R~\rho\sigma}=\delta^{\mu\nu}_{\rho\sigma}/2$ and $k=1/2$.
To this point, the tensor $\mathcal{P}^{\mu\nu}_{R~\rho\sigma}$ could be
regarded as the generalization of $\mathcal{P}^{\mu\nu}_{(1)\rho\sigma}$
in the context of the generic gravity theories described by the Lagrangian
(\ref{CalLRiem}).

In parallel with the situation of the Lovelock gravity theory, motivated by
the Noether potential $K^{\mu\nu}_R$ via the covariant phase space method
\cite{LeeWald,IyWald,WalZo}, that is,
\be
K^{\mu\nu}_R=P^{\mu\nu\rho\sigma}_{R}\nabla_\rho\xi_\sigma
-2\xi_\sigma\nabla_\rho P^{\mu\nu\rho\sigma}_{R}
\, , \label{IWalPot}
\ee
we make use of the tensor $\mathcal{P}^{\mu\nu\rho\sigma}_R$ to replace
$P^{\mu\nu\rho\sigma}_R$ in the above equation to
introduce a similar superpotential $\mathcal{K}_R^{\mu\nu}$, which is
given by
\be
\mathcal{K}_R^{\mu\nu}=
\mathcal{P}^{\mu\nu\rho\sigma}_{R}\nabla_\rho\xi_\sigma
-2\xi_\sigma\nabla_\rho \mathcal{P}^{\mu\nu\rho\sigma}_{R}
\, . \label{CalKRdef}
\ee
Alternately, it is convenient to express $\mathcal{K}_R^{\mu\nu}$ as
\be
\mathcal{K}_R^{\mu\nu}=K_R^{\mu\nu}-\frac{6k}{(D-3)\ell}
R^{[\rho\sigma}_{\rho\sigma}\nabla^\mu\xi^{\nu]}
+k(D-4)\nabla^{[\mu}\xi^{\nu]}
\, . \label{CalKRdef2}
\ee
In some sense, the superpotential $\mathcal{K}_R^{\mu\nu}$ can be called as
Komar-like potential if it is interpreted as the generalization of the
ordinary Komar potential $P^{\mu\nu}_{(0)\rho\sigma}\nabla^\rho\xi^\sigma$
associated with the Einstein gravity theory. According to Eq. (\ref{delP1xi}),
the perturbation of $\mathcal{K}_R^{\mu\nu}$ on the AdS reference background
leads to
\bea
\delta\mathcal{K}_R^{\mu\nu}&=&\bar{Q}_R^{\mu\nu}
-\frac{k}{(D-3)\ell}\bar{\nabla}_\gamma \bar{U}^{\gamma\mu\nu}_{(1)}\, , \nn \\
\bar{Q}_R^{\mu\nu}&=&\big(\delta P^{\mu\nu}_{R~\rho\sigma}\big)
\bar{\nabla}^\rho\bar{\xi}^\sigma
-2\bar{\xi}^\sigma\bar{\nabla}^\rho
\delta P^{\mu\nu}_{R~\rho\sigma}+2k\bar{Q}_{EH}^{\mu\nu} \nn \\
&=&\delta K^{\mu\nu}_{R}
+\frac{1}{2}h\bar{P}^{\mu\nu\rho\sigma}_{R}
\bar{\nabla}_\rho\bar{\xi}_\sigma
-2\bar{\xi}^{[\mu}\bar{P}^{\nu]\lambda\rho\sigma}_{R}
\bar{\nabla}_\sigma h_{\rho\lambda}
\, . \label{PerCalKR}
\eea
As before, by following the works \cite{KimKY,JJPEPJC}, one can verify that
the 2-form $\bar{Q}_R^{\mu\nu}$ is the (off-shell) ADT potential on a fixed
AdS background, and it also coincides with the results in \cite{AmGor}.
What is more, in terms of Eq. (\ref{PRgeform}), one finds
that $\delta P^{\mu\nu}_{R~\rho\sigma}$ could be generally expressed as
\be
\delta P^{\mu\nu}_{R~\rho\sigma}=
\lambda_1\delta R^{\mu\nu}_{~~\rho\sigma}
+\lambda_2\delta R^{[\mu}_{[\rho} \delta^{\nu]}_{\sigma]}
+\lambda_3\delta R \delta^{\mu\nu}_{\rho\sigma}
+\lambda_4 h^{[\mu}_{[\rho} \delta^{\nu]}_{\sigma]}
\, , \label{delPR}
\ee
where $\lambda_i$'s are constant parameters and the concrete expressions for
$\delta R^{\mu\nu}_{~~\rho\sigma}$, $\delta R^{\mu}_{\rho}$ and $\delta R$
were given in the appendix of Ref. \cite{JJPEPJC}. For instance, when
$\lambda_1=1$, $\lambda_2=-4$, $\lambda_3=1/2$ and $\lambda_4=0$, one
obtains
\be
\delta \mathcal{P}^{\mu\nu}_{(1)\rho\sigma}=
\delta R^{\mu\nu}_{~~\rho\sigma}
-4\delta R^{[\mu}_{[\rho} \delta^{\nu]}_{\sigma]}
+\frac{1}{2}\delta R \delta^{\mu\nu}_{\rho\sigma}
\, . \label{delP1def}
\ee

In terms of $\delta\mathcal{K}_R^{\mu\nu}$, when the dimension $D>3$,
a formula for the conserved charges of the gravity theory described by
the Lagrangian (\ref{CalLRiem}) can be proposed as a Komar-like integral
\be
\mathcal{Q}_{Riem}=\frac{1}{8\pi}\int_{\partial\Sigma}
\delta\mathcal{K}^{\mu\nu}_{R} d\Sigma_{\mu\nu}
\, , \label{CCofLRiem}
\ee
which is able to be completely determined by Eqs. (\ref{delPR}) and
(\ref{delP1def}). Apart from $\mathcal{P}_R^{\mu\nu\rho\sigma}$, another
fourth-rank tensor $\tilde{\mathcal{P}}_R^{\mu\nu\rho\sigma}$,
given by\footnote{If $\tilde{\mathcal{P}}^{\mu\nu\rho\sigma}_{R}$ is not
required to exhibit all the algebraic symmetry properties for the Riemann
tensor, it can be alternatively defined as
$
\tilde{\mathcal{P}}^{\mu\nu}_{R~\rho\sigma}\rightarrow
P^{\mu\nu}_{R~\alpha\beta}
\mathcal{P}^{\alpha\beta}_{(1)\rho\sigma}
-2(D-3)\ell\big(P^{\mu\nu}_{R~\rho\sigma}
-k\delta^{\mu\nu}_{\rho\sigma}\big)$. Unlike
$\mathcal{P}^{\mu\nu\rho\sigma}_{R}$, here
$\tilde{\mathcal{P}}^{\mu\nu\rho\sigma}_{R}=
\mathcal{P}^{\mu\nu\rho\sigma}$ for the Einstein gravity.}
\be
\tilde{\mathcal{P}}^{\mu\nu}_{R~\rho\sigma}=
\frac{1}{2}\Big(P^{\mu\nu}_{R~\alpha\beta}
\mathcal{P}^{\alpha\beta}_{(1)\rho\sigma}+
\mathcal{P}^{\mu\nu}_{(1)\alpha\beta}
P^{\alpha\beta}_{R~\rho\sigma}\Big)
-2(D-3)\ell\big(P^{\mu\nu}_{R~\rho\sigma}-k\delta^{\mu\nu}_{\rho\sigma}\big)
\, , \label{TCaPRdef}
\ee
can also be adopted to define the conserved charges. To see this clearly,
by introducing the 2-form
\be
\tilde{\mathcal{K}}_R^{\mu\nu}
=\tilde{\mathcal{P}}^{\mu\nu\rho\sigma}_R
\nabla_\rho\xi_\sigma+\frac{1}{(D-3)\ell}
\xi_\sigma\nabla_\rho\tilde{\mathcal{P}}^{\mu\nu\rho\sigma}_R
\, , \label{TCaKRdef}
\ee
whose perturbation on the AdS background yields
\be
\delta\tilde{\mathcal{K}}_R^{\mu\nu}=-2(D-3)\ell \bar{Q}_R^{\mu\nu}
+2k\bar{\nabla}_\gamma \bar{U}^{\gamma\mu\nu}_{(1)}
\, , \label{PerTCaKR}
\ee
the formula (\ref{CCofLRiem}) is rewritten as
\be
\mathcal{Q}_{Riem}=-\frac{1}{16\pi(D-3)\ell}\int_{\partial\Sigma}
\delta\tilde{\mathcal{K}}^{\mu\nu}_{R} d\Sigma_{\mu\nu}
\, . \label{CCofLRiem2}
\ee
As an example to demonstrate the conserved quantity $\mathcal{Q}_{Riem}$,
substituting
\be
P^{\mu\nu}_{R~\rho\sigma}=P^{\mu\nu}_{(i)\rho\sigma}\, , \quad
k=\lambda_{(i)}=\Big(\frac{\ell}{2}\Big)^i\frac{(D-2)!}{(D-2i-2)!}
\,  \nn
\ee
into the formula (\ref{CCofLRiem}) or (\ref{CCofLRiem2}), one obtains the
conserved charge $\mathcal{Q}_{(i)}$ for the Lovelock-type Lagrangian
(\ref{LovGLag}), which has been presented in Eq. (\ref{CCinCalK}).

A remark is in order here. By contrast with the conventional ADT formalism,
our definition for the conserved charges at least has the following merits.
First, calculations on the potential become more operable. It straightforwardly
arises from the perturbation of $\mathcal{K}_R^{\mu\nu}$ on the AdS
background. Second, because of the vanishing of the tensor
$\mathcal{P}_R^{\mu\nu\rho\sigma}$ on the background spacetimes, apart from
the variation of $\mathcal{K}_R^{\mu\nu}$, the 2-form $\mathcal{K}_R^{\mu\nu}$
itself can be directly adopted to enter into the surface integral.
As a consequence, all the computations with respect to the perturbation of
the gravitational field are avoidable, extremely simplifying the formula.
Third, the potential $\mathcal{K}_R^{\mu\nu}$ takes a closer form to that
through the well-known covariant phase space method. Thus, the formulation
in the present work gives assistance to built a connection between the
ADT approach and the covariant phase space method. Actually, in comparison
with the Iyer-Wald potential given by the covariant phase space method,
the quantity $\delta\big(\mathcal{K}_R^{\mu\nu}-K_R^{\mu\nu}\big)$ just
compensates the contribution from the surface term, rendering
$\delta\mathcal{K}_R^{\mu\nu}$ equivalent to the Iyer-Wald potential.
Furthermore, with the guidance of such an equivalence, we arrive at the
conclusion that the ADT potential is equivalent with the one via the
covariant phase space method in the context of the generic gravities
endowed with the Lagrangian (\ref{CalLRiem}). However, an obvious demerit
of the formula (\ref{CCofLRiem}) for the conserved charges is that it does
not incorporate the contribution for the matter fields, which is of great
importance, particulary to the conserved charges of G\"{o}del-type black
holes \cite{PJJGCC}. This deserves to be dealt with in future.

\section{Summary}\label{five}

In the present paper, within the framework of the Lovelock gravity theory, we propose
a new rank-four divergenceless tensor $\mathcal{P}^{\mu\nu\rho\sigma}_{(i)}$
given in Eq. (\ref{MCPidef}). This tensor is dependent of the Riemann curvature
tensor and preserves its all algebraic symmetries. In terms of the tensor
$\mathcal{P}^{\mu\nu\rho\sigma}_{(i)}$, a 2-form $\mathcal{K}^{\mu\nu}_{(i)}$
associated with an arbitrary Killing vector $\xi^\mu$ is constructed. The perturbation of
$\mathcal{K}^{\mu\nu}_{(i+1)}$ on the AdS background serves as an ideal
superpotential in the definition for the conserved charges of the Lovelock
gravity theories in asymptotically AdS spacetimes, and it further give rises
to the formula (\ref{CCinCalK}). Subsequently, inspired by the tensor
$\mathcal{P}^{\mu\nu\rho\sigma}_{(i)}$, a fourth-rank tensor
$\mathcal{P}^{\mu\nu\rho\sigma}_{R}$
(or $\tilde{\mathcal{P}}^{\mu\nu\rho\sigma}_{R}$) is constructed in the
context of the general diffeomorphism invariant theories of gravity
with the Lagrangian (\ref{CalLRiem}).
In parallel, with the help of this tensor, as well as its perturbation upon
the AdS background, we have put forward the definition for the conserved
charges $\mathcal{Q}_{Riem}$ in Eq. (\ref{CCofLRiem}) or (\ref{CCofLRiem2}),
which is applicable to an arbitrary gravity theory depending merely on the
curvature tensor. What is more, we have clarified that all the
newly-constructed formulas of the conserved charges are equivalent with the
ones via the ADT method and the covariant phase space approach.

\section*{Acknowledgments}

We would like to thank the anonymous referees for their valuable suggestions
and comments. This work was supported by the Natural Science Foundation of
China under Grant Nos. 11865006 and 11505036. It was also partially supported
by the Technology Department of Guizhou province Fund
under Grant Nos. [2018]5769.

\end{document}